\documentclass[3p,times,twocolumn]{elsarticle}

\usepackage{ecrc}

\volume{00}

\firstpage{1}

\journalname{Nuclear Physics B Proceedings Supplement}

\runauth{}

\jid{nuphbp}

\jnltitlelogo{Nuclear Physics B Proceedings Supplement}

\usepackage{amssymb}

 \biboptions{square}
 \biboptions{sort&compress}

\usepackage[figuresright]{rotating}

\begin{document}

\begin{frontmatter}



\dochead{}

\title{Exclusive production of $\chi_{c}(0^{+})$ meson and its measurement in the $\pi^{+}\pi^{-}$ channel}


\author[label1]{P. Lebiedowicz}
\ead{piotr.lebiedowicz@ifj.edu.pl}

\author[label2]{R. Pasechnik}
\ead{roman.pasechnik@fysast.uu.se}

\author[label1,label3]{A. Szczurek}
\ead{antoni.szczurek@ifj.edu.pl}

\address[label1]{Institute of Nuclear Physics PAN, PL-31-342 Cracow, Poland}
\address[label2]{Department of Physics and Astronomy, Uppsala University, Box 516, SE-751 20 Uppsala, Sweden}
\address[label3]{University of Rzesz\'ow, PL-35-959 Rzesz\'ow, Poland}

\begin{abstract}
We report on the results of a theoretical study
of the central exclusive production of scalar
$\chi_{c}(0^{+})$ meson via $\chi_{c0} \to \pi^{+}\pi^{-}$ decay
in high-energy hadron collisions at the RHIC, Tevatron and LHC.
The corresponding amplitude for exclusive
double-diffractive $\chi_{c0}$ meson production
was obtained within the $k_{t}$-factorization approach
including virtualities of active gluons
and the cross section is calculated with
unintegrated gluon distribution functions (UGDFs)
known from the literature.
The four-body $p p \to p p \pi^+ \pi^-$ reaction constitutes
an irreducible background to the exclusive
$\chi_{c0}$ meson production.
We include the absorption effects
due to proton-proton interaction and
pion-pion rescattering.
Several differential distributions
for $pp(\bar{p}) \to pp(\bar{p})\chi_{c0}$
process, including the absorptive corrections, were calculated.
The influence of kinematical cuts on the signal-to-background ratio
is investigated. 
\end{abstract}


\end{frontmatter}


\section{Introduction}
The mechanism of exclusive production of mesons at high energies
became recently a very active field of research (see \cite{ACF10} and refs. therein).
These reactions $p p \to pMp$,
where $M=\sigma, \rho^{0}, f_{0}(980), \phi, f_{2}(1270), f_{0}(1500)$, $\chi_{c0}$,
provide a valuable tool to investigate in detail the properties of resonance states at high energies.
The recent works concentrated on the production of
$\chi_c$ mesons (see e.g. Refs. \cite{PST_chic0, PST_chic1, PST_chic2, LKRS10})
where the QCD mechanism is similar to the
exclusive production of the Higgs boson. Furthermore, the
$\chi_{c(0,2)}$ states are expected to annihilate via two-gluon
processes into light mesons and may, therefore, allow the study of
glueball production dynamics.
The two-pion background to exclusive production of $f_{0}(1500)$ \cite{SL09} 
and $\chi_{c0}$ \cite{LPS11} mesons was already discussed.
In Ref. \cite{LKRS11} a new perturbative mechanism of the $\pi\pi$ production was discussed.
Due to reasons explained in Ref. \cite{LKRS11} this mechanism gives 
relatively small contribution in the $\chi_{c0}$ invariant mass region.

The CDF Collaboration has measured the cross section of CEP $\chi_c$ mesons 
in proton-antiproton collisions at the Tevatron \cite{CDF_chic}.
In this experiment $\chi_c$ mesons are identified via decay to the $J/\psi + \gamma$
with $J/\psi \to \mu^{+}\mu^{-}$ channel. The experimental invariant
mass resolution was not sufficient to distinguish between scalar,
axial and tensor $\chi_c$. While the branching fractions to this
channel for axial and tensor mesons are large \cite{PDG}
($\mathcal{B} = (34.4 \pm 1.5)\%$ and $\mathcal{B} = (19.5 \pm 0.8)\%$, respectively)
the branching fraction for the scalar meson is very small 
$\mathcal{B} = (1.16 \pm 0.08 )\%$ \cite{PDG}. 
On the other hand, the cross section
for exclusive $\chi_{c0}$ production obtained within the
$k_{t}$-factorization is much bigger than that for $\chi_{c1}$ and
$\chi_{c2}$. As a consequence, all $\chi_{c}$ mesons give similar
contributions \cite{PST_chic2} to the $J/\psi + \gamma$ decay
channel. Clearly, the measurement via decay to the $J/\psi + \gamma$
channel cannot provide cross section for different $\chi_c$.

The $\chi_{c0}$ meson decays
into several two-body channels (e.g. $\pi \pi$, $K^{+} K^{-}$, $p \bar{p}$)
or four-body hadronic modes (e.g. $\pi^{+} \pi^{-} \pi^{+} \pi^{-}$,
$\pi^{+} \pi^{-} K^{+} K^{-}$).
We have analyzed a possibility to measure $\chi_{c0}$ via its decay into
$\pi^+ \pi^-$ channel \cite{LPS11}.
The branching fraction $\mathcal{B}( \chi_{c0} \to \pi^{+} \pi^{-}) = (0.56 \pm 0.03)\%$ is large,
the axial $\chi_{c1}$ does not decay to the $\pi \pi$ channel 
and $\mathcal{B}( \chi_{c2} \to \pi^{+} \pi^{-}) = (0.16 \pm 0.01)\%$ is smaller \cite{PDG}.
In addition a much smaller cross section for $\chi_{c2}$ production means
that in practise only $\chi_{c0}$ will contribute to the signal.
The advantage of this channel is that the $\pi^+ \pi^-$
continuum has been studied recently \cite{LS10,LPS11} and is relatively well known.

\section{Signal and background amplitudes}
\label{section:Signal and background amplitudes}

\begin{figure}[!ht]    
\begin{center}
 \includegraphics[width=0.25\textwidth]{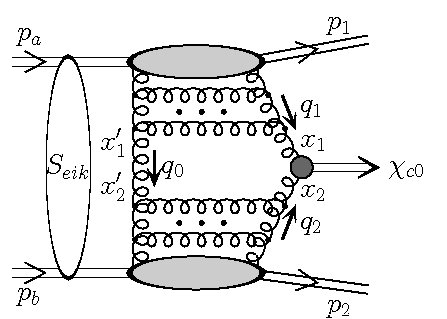}
\end{center}
   \caption{\label{fig:fig1}
   \small  The QCD mechanism of exclusive diffractive production of $\chi_{c0}$ meson
   including the absorptive correction.}
\end{figure}
The QCD mechanism for the diffractive production of heavy central
system has been proposed by Khoze, Martin and Ryskin (KMR) and
developed in collaboration with Kaidalov and Stirling for Higgs
production (see e.g. Refs.~\cite{KMR}).
In the framework of this
approach the amplitude of the exclusive $pp\to pp \chi_{c0}$ process
is described by the diagram shown in Fig.~\ref{fig:fig1}, 
where the hard subprocess $g^{*}g^{*} \to \chi_{c0}$ is initiated by the
fusion of two off-shell gluons and the soft part is represented in
terms of the off-diagonal unintegrated gluon distribution functions (UGDFs).
The formalism used to calculate the exclusive $\chi_{c0}$ meson
production is explained in detail elsewhere \cite{PST_chic0}.

\begin{figure}[!ht]
\begin{center}
\includegraphics[width=0.2\textwidth]{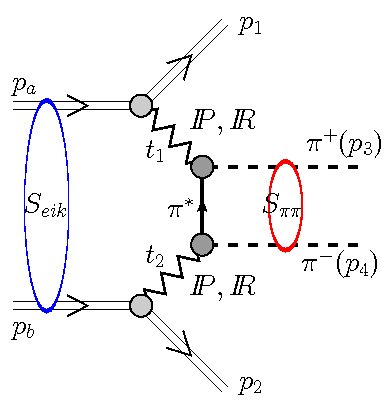}
\includegraphics[width=0.2\textwidth]{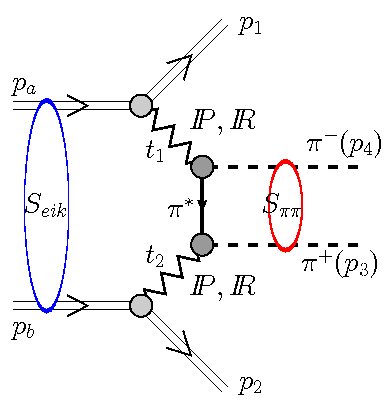}
\end{center}
  \caption{\label{fig:central_double_diffraction_diagrams_fsi}
  \small
The double-diffractive mechanism of exclusive production of
$\pi^{+}\pi^{-}$ pairs including the pion and Regge exchanges,
the absorptive corrections due to proton-proton interactions as well as
pion-pion rescattering.}
\end{figure}

The dominant mechanism of the exclusive production of
$\pi^{+}\pi^{-}$ pairs at high energies is sketched in
Fig.~\ref{fig:central_double_diffraction_diagrams_fsi}.
The expected non-resonant background can be modeled using a
"non-perturbative" framework where the pion pair is produced by 
Pomeron-Pomeron fusion
with an intermediate off-shell pion/Reggeon exchanged between the
final-state particle pairs (see \cite{LS10,LS11,LPS11} for details).
In calculations of the amplitude we follow the general rules of
Pumplin and Henyey \cite{PH76} used recently in Ref.~\cite{LS10}
where a first estimate of the differential cross sections
for the $\pi^+ \pi^-$ pairs production at the LHC energies has been presented.
The Regge parametrization of the $\pi^{\pm} p \to \pi^{\pm} p$ and $\pi^+ \pi^- \to \pi^+ \pi^-$
scattering amplitude includes both Pomeron
as well as $f$ and $\rho$ Reggeon exchanges
with the parameters taken from
the Donnachie-Landshoff analysis \cite{DL92} of the total cross sections.
The Regge-type interaction applies at higher energies and 
at low energies should be switched off (see \cite{LS10, LS11, LPS11}).
In Ref. \cite{LPS11} we propose to use a generalized propagator. 
The form factors correct for the off-shellness of the intermediate pions
are parametrized as 
$F_{\pi}(\hat{t}/\hat{u})=
\exp\left(\frac{\hat{t}/\hat{u}-m_{\pi}^{2}}{\Lambda^{2}_{off}}\right)$,
where the parameter $\Lambda_{off}$ is obtained from fit 
to the experimental data \cite{ABCDHW90} (see \cite{LPS11}).

\section{Results}
\label{section:Results}
\begin{figure}[!ht]
\includegraphics[width = 0.23\textwidth]{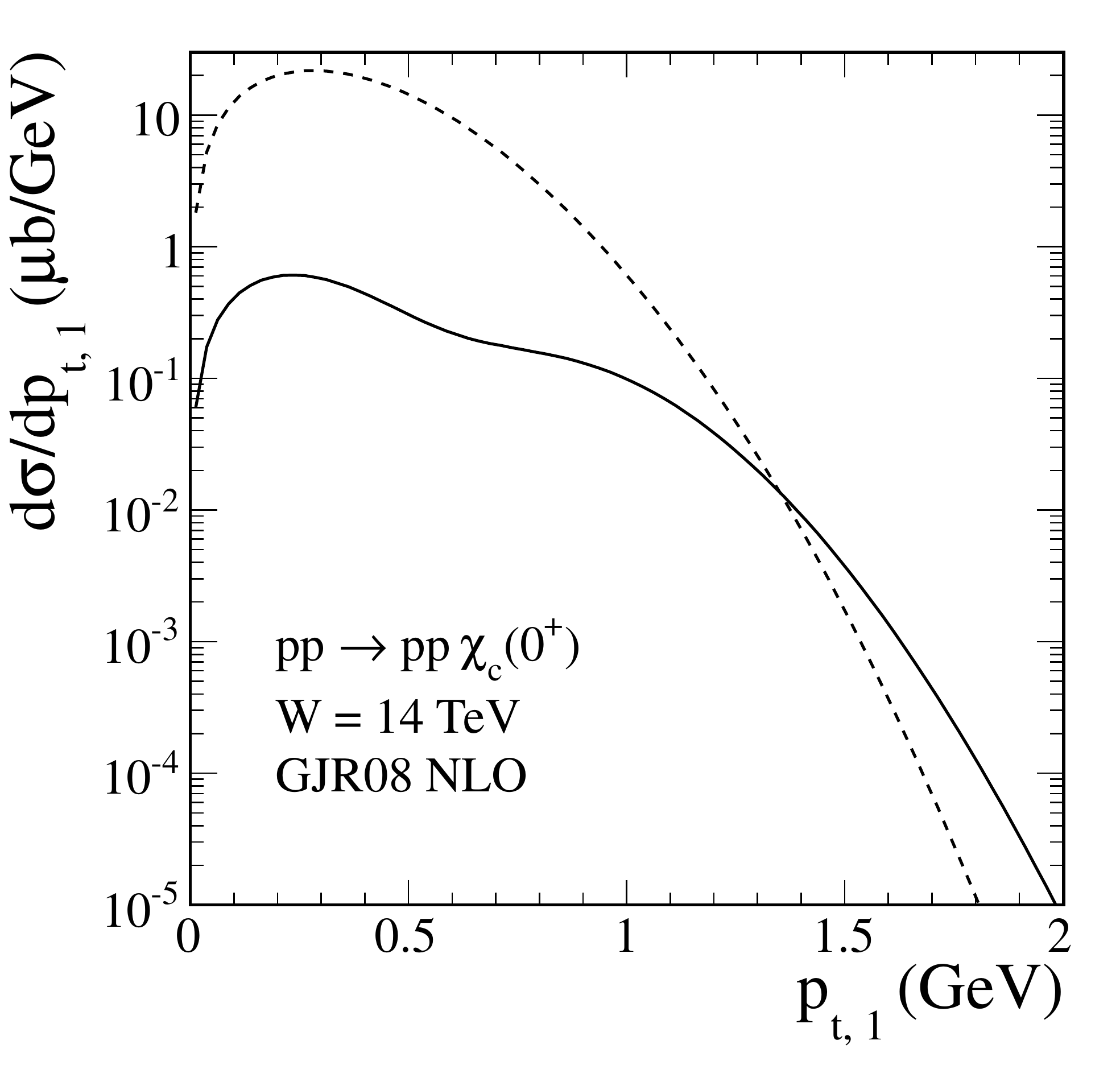}
\includegraphics[width = 0.23\textwidth]{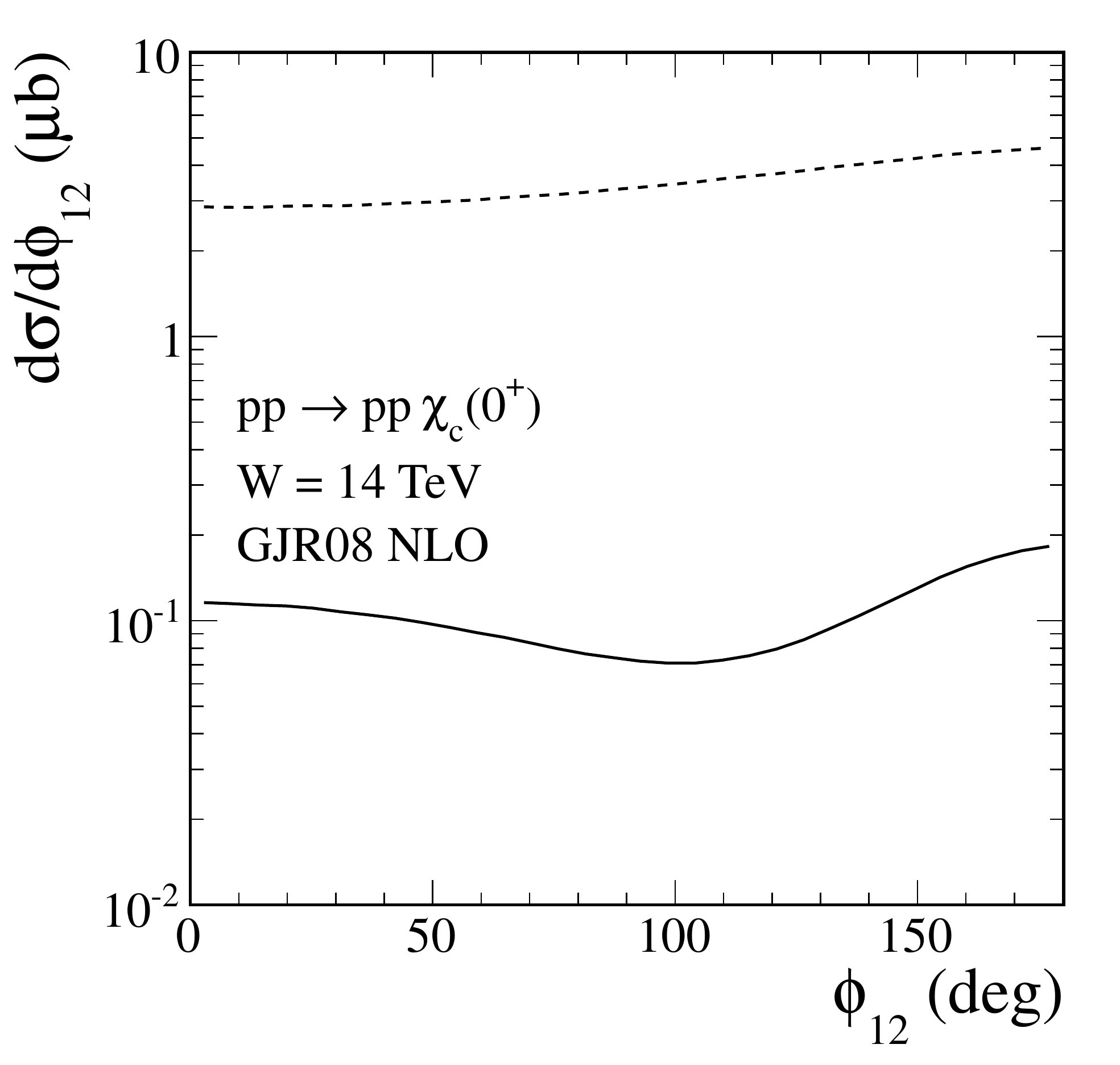}\\
\includegraphics[width = 0.23\textwidth]{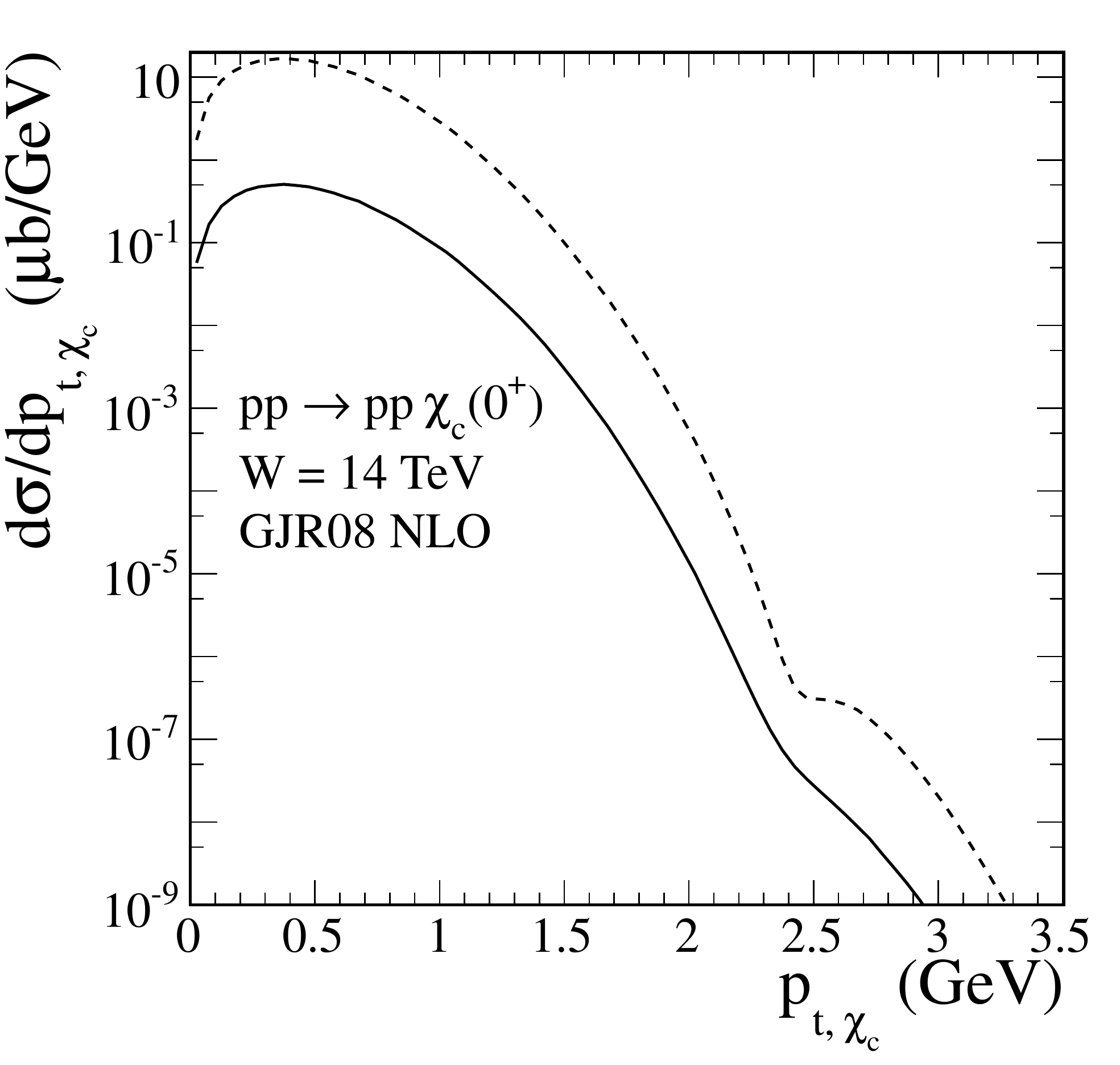}
\includegraphics[width = 0.23\textwidth]{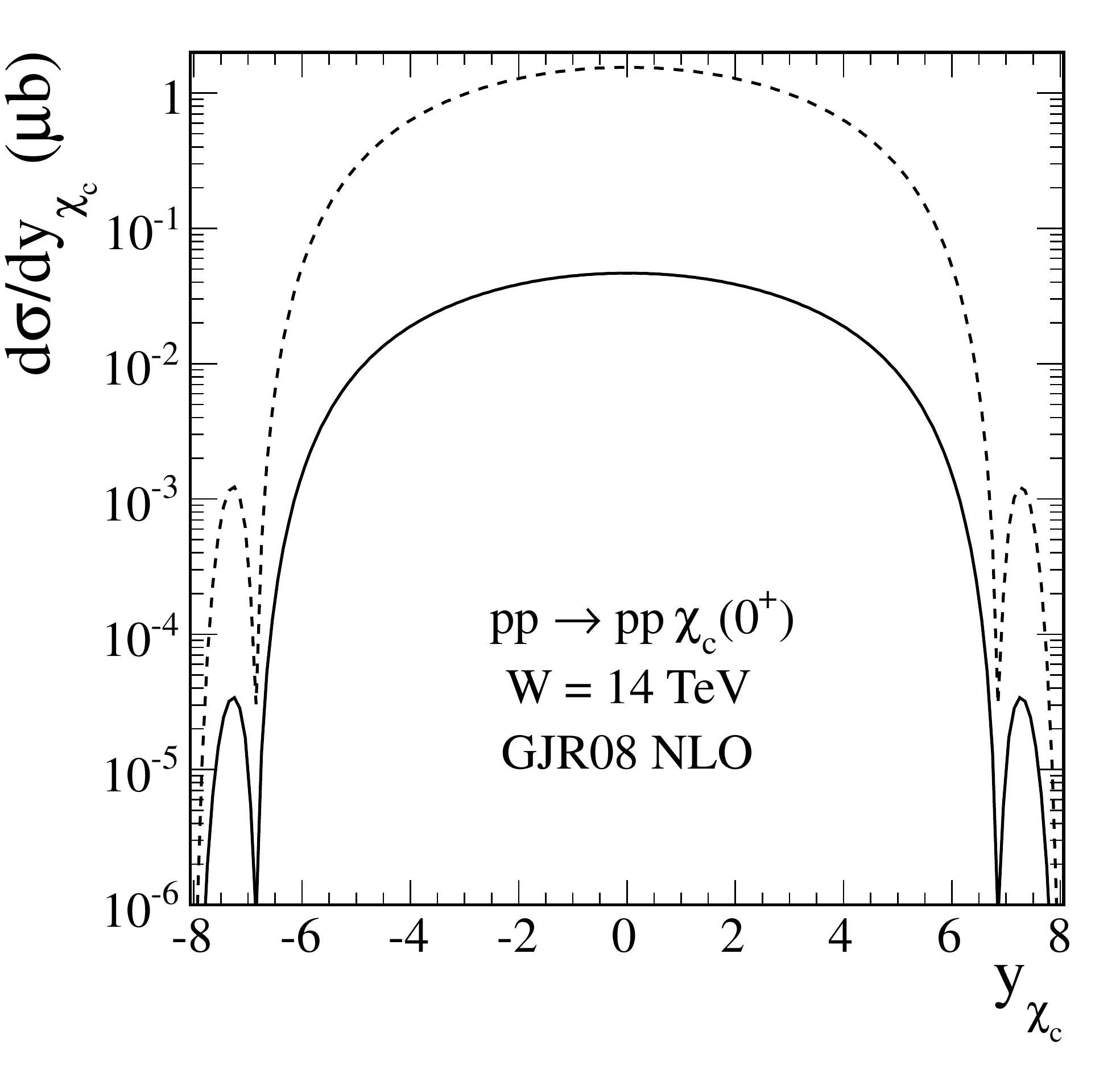}
  \caption{\label{fig:diff_comp}
  \small
Differential cross sections for the $pp \to pp \chi_{c0}$ reaction
at $\sqrt{s}$ = 14 TeV without (dashed line) and with (solid line)
absorption effects.}

\end{figure}

We first show (Fig.~\ref{fig:diff_comp}) 
the differential cross sections of $\chi_{c0}$ CEP at $\sqrt{s}$ = 14 TeV 
without (dashed line) and with (solid line) absorptive corrections.
These calculations were done with GJR NLO \cite{GJR} collinear gluon
distribution, to generate the KMR UGDFs,
which allows to use low values of the internal gluon transverse momenta
$Q_{t}^2 \geq Q_{cut}^2 = 0.5$ GeV$^2$. The bigger the value of the
cut-off parameter, the smaller the cross section (see
Ref.~\cite{PST_chic0}). In the calculations we take the value of the
hard scale to be $\mu^2 = M^{2}$. The smaller $\mu^2$, the bigger
the cross section \cite{PST_chic0}. 
In all cases the absorption effects lead to a damping of the cross section.
In most cases the shape is almost unchanged.
Exception is the distribution in proton transverse momentum where
the absorption effects lead to a damping of the cross section at small
proton $p_{t}$ and an enhancement of the cross section at large proton $p_{t}$. 
In relative azimuthal angle distribution 
we observe a dip at $\phi_{12}\sim \pi/2$. 
Transverse momentum distribution of $\chi_{c0}$ shows a small minimum at $p_{t} \sim$ 2.5 GeV. 
The main reason of its appearance is the functional dependence of matrix elements on its
arguments \cite{PST_chic0}.

\begin{figure}[!ht]
\includegraphics[width = 0.23\textwidth]{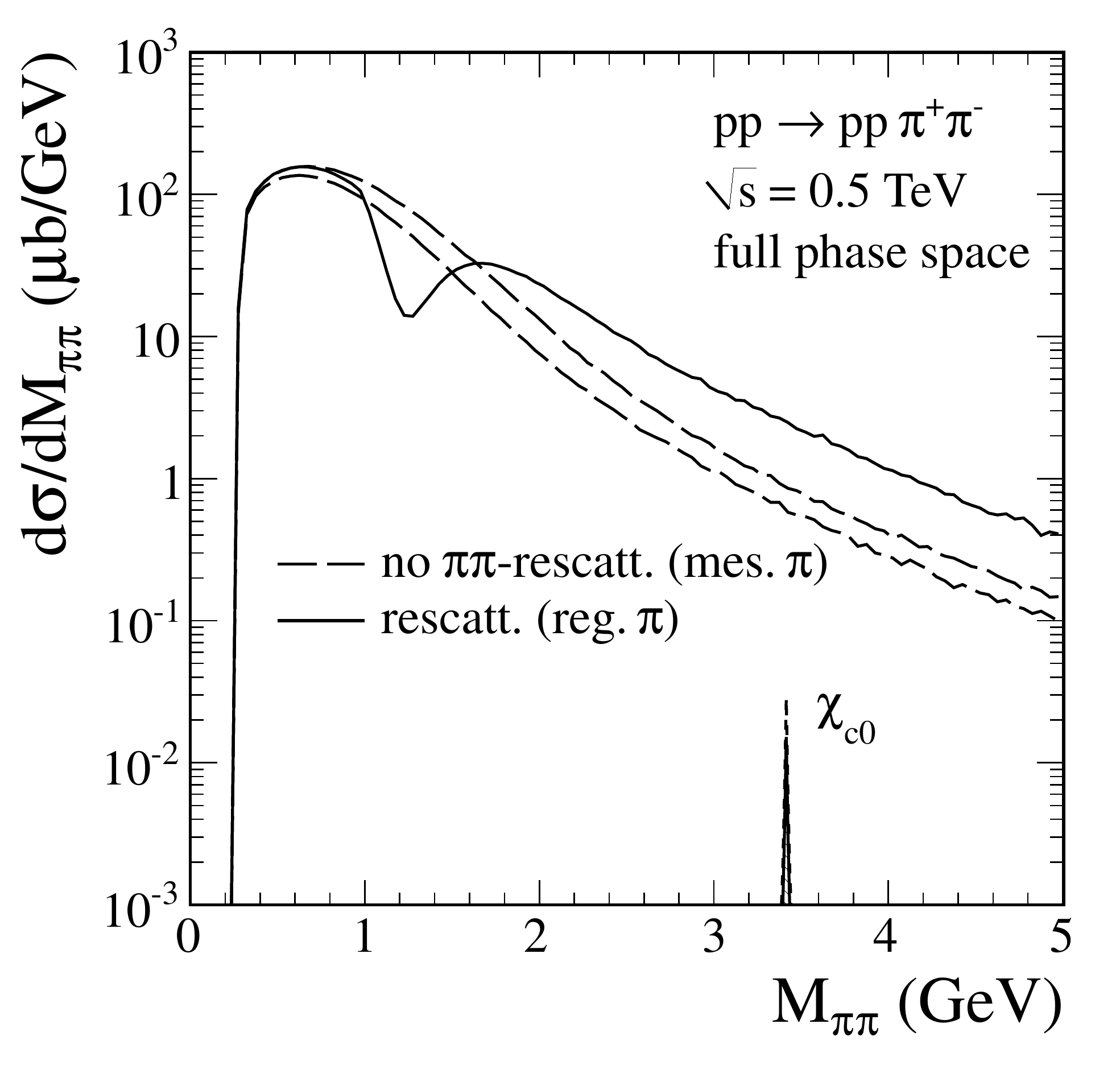}
\includegraphics[width = 0.23\textwidth]{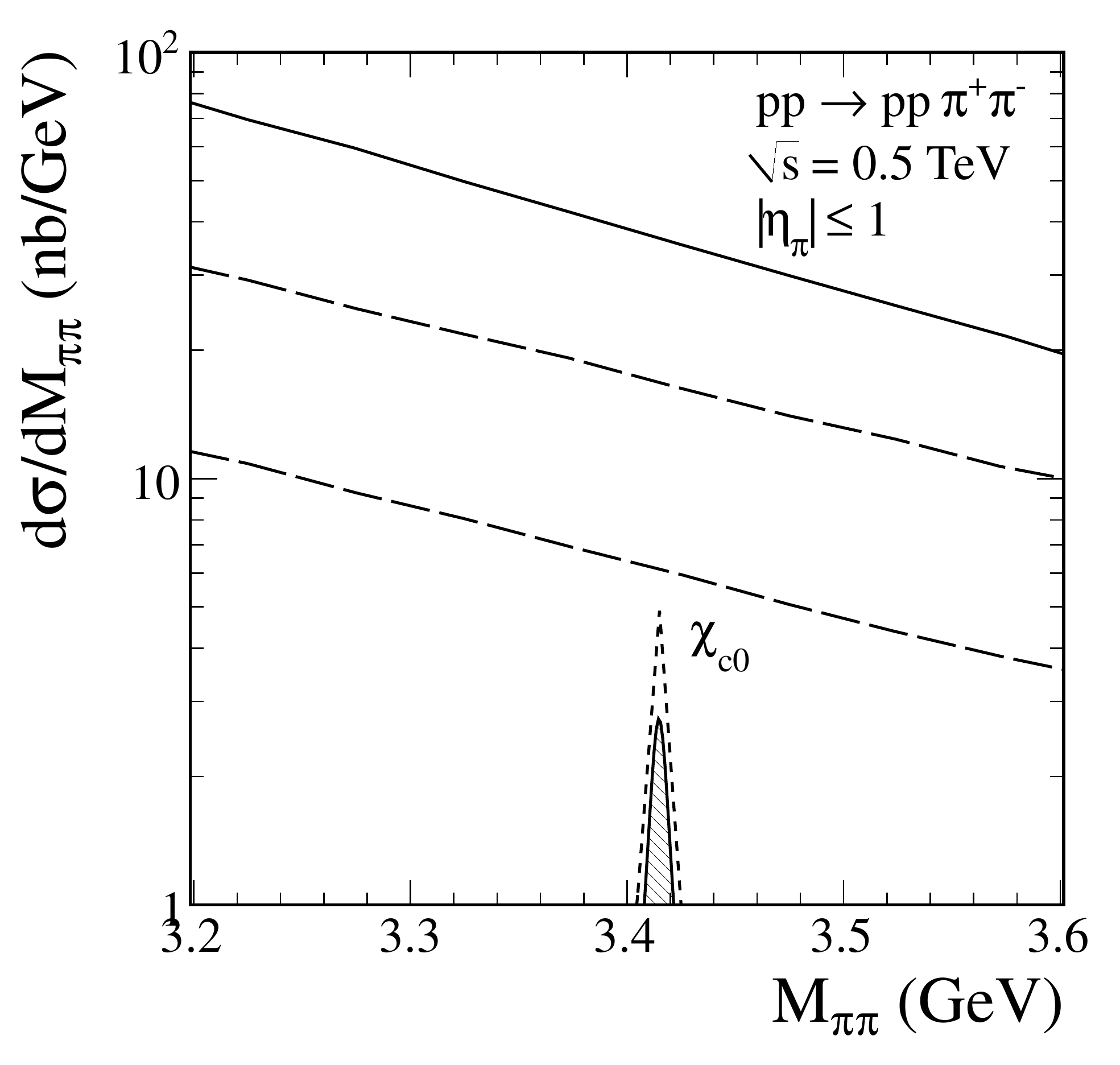}
\includegraphics[width = 0.23\textwidth]{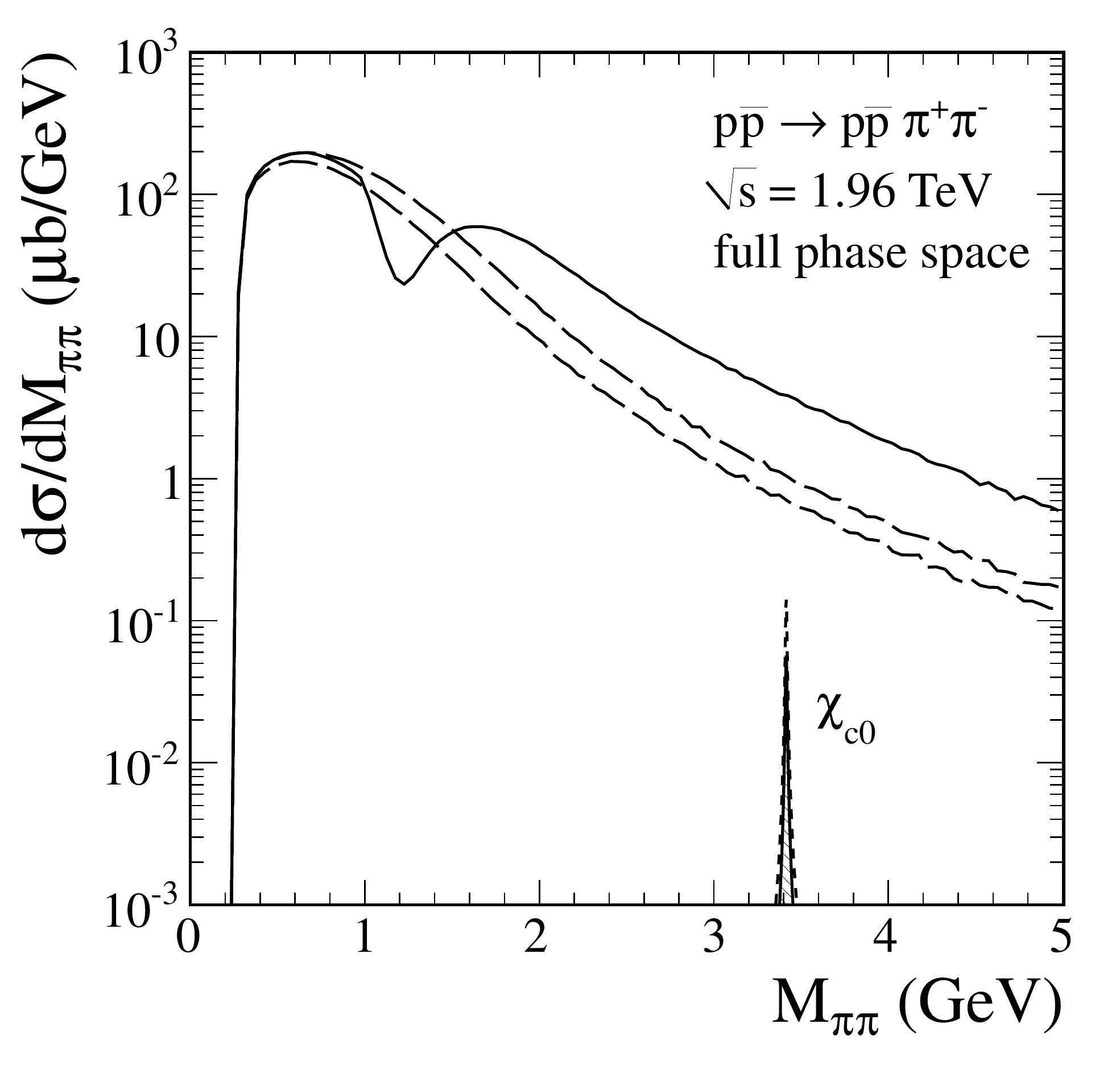}
\includegraphics[width = 0.23\textwidth]{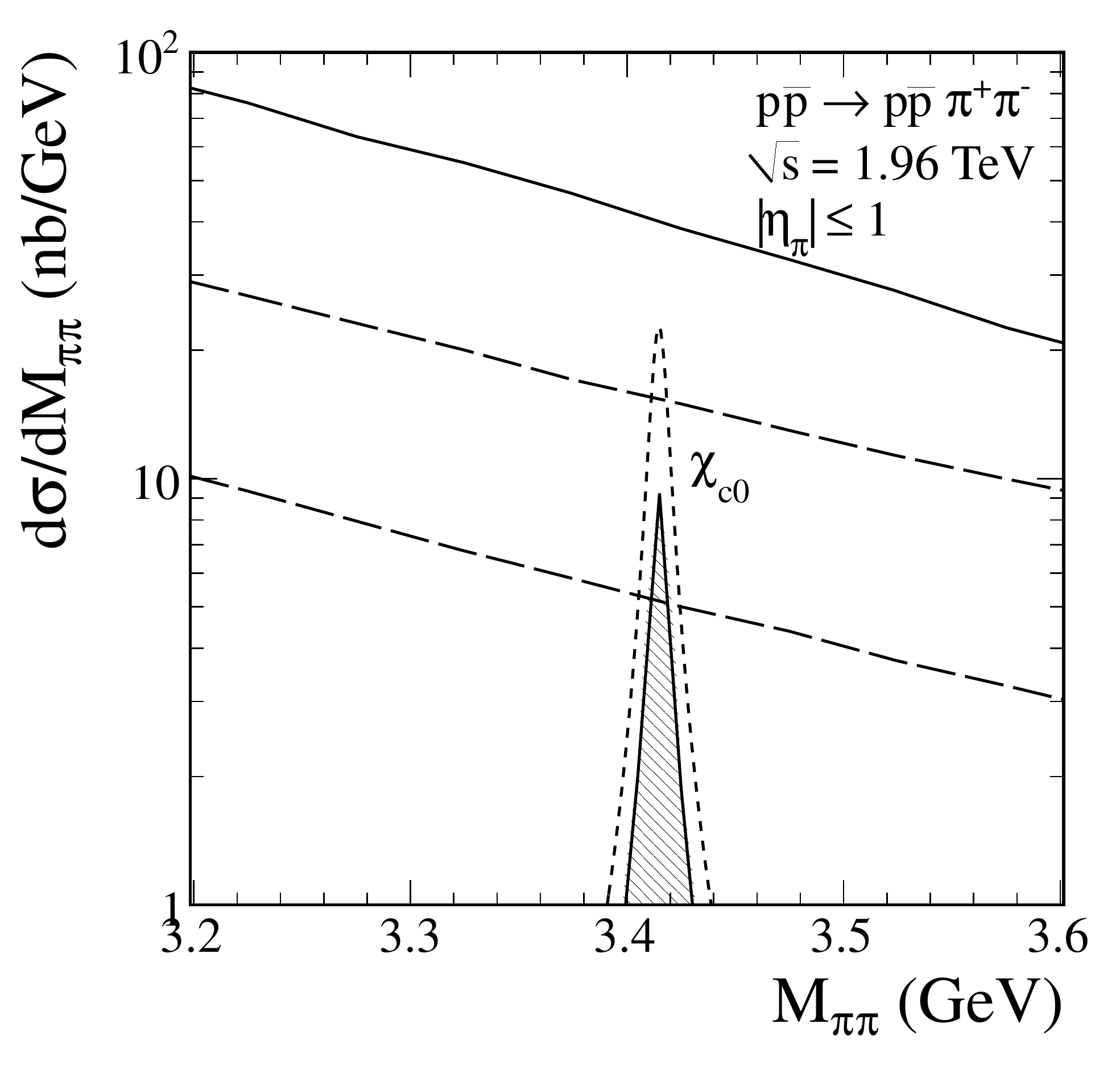}
\includegraphics[width = 0.23\textwidth]{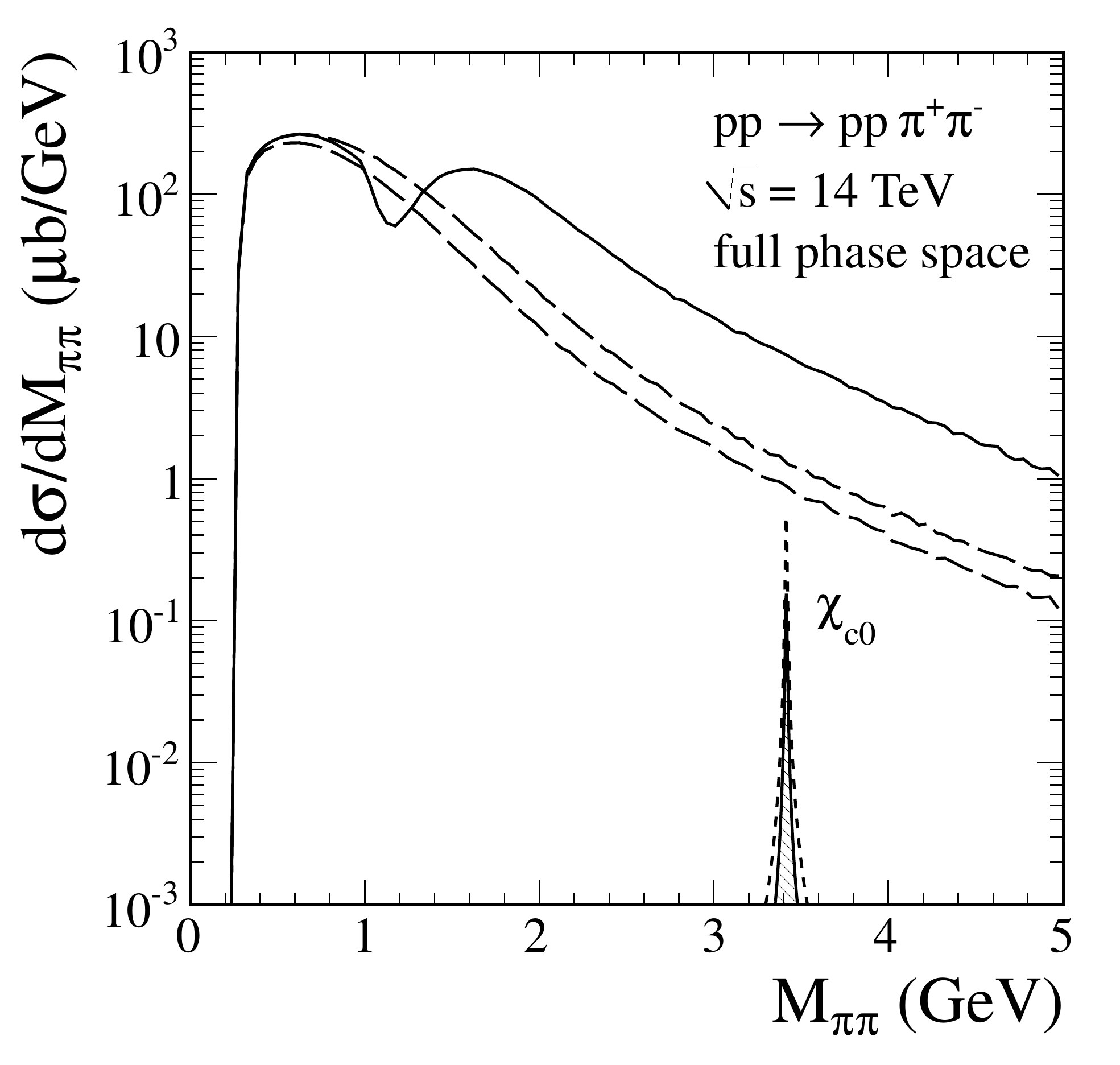}
\includegraphics[width = 0.23\textwidth]{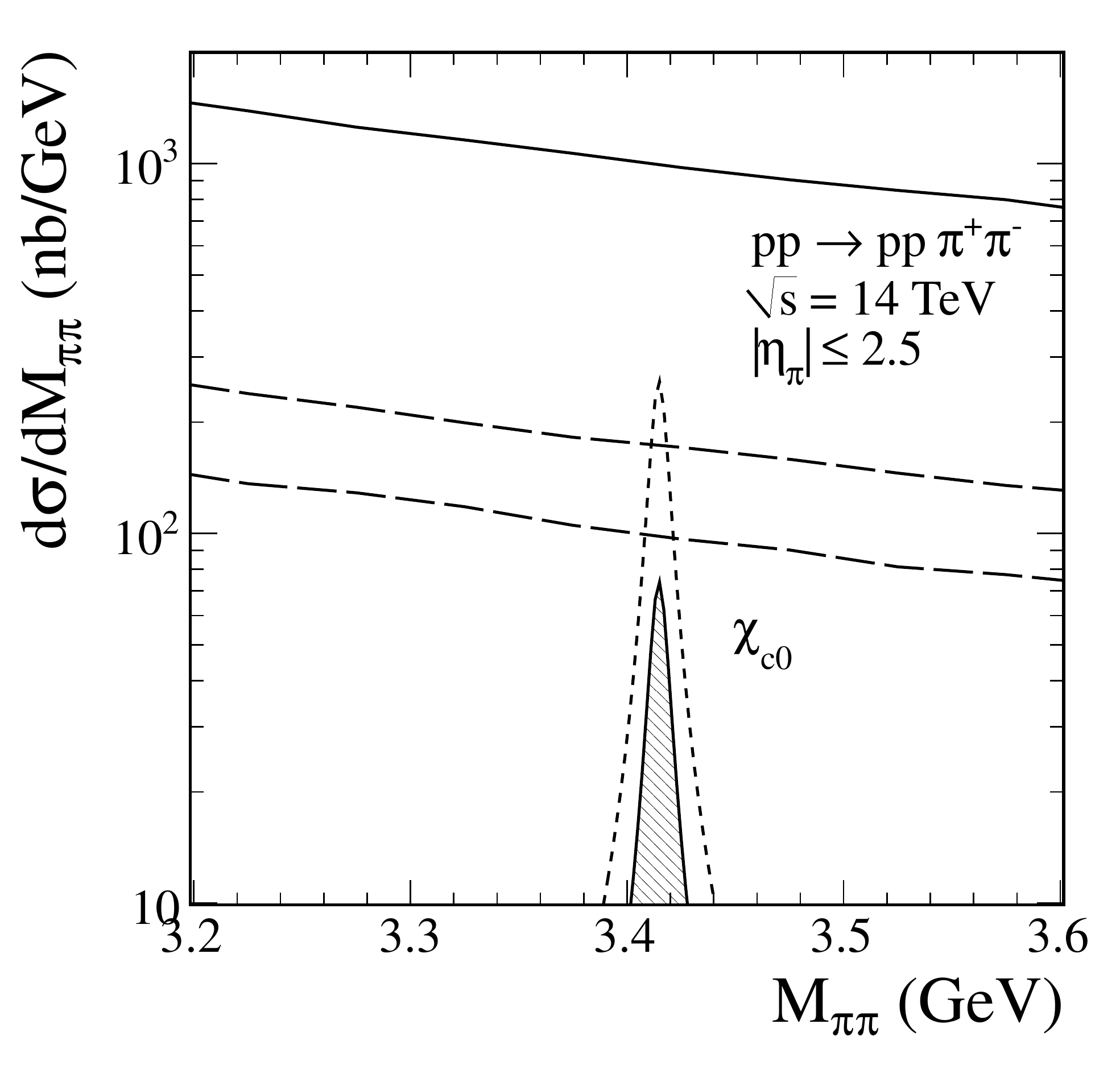}
  \caption{\label{fig:dsig_dmpipi}
  \small
The $\pi^{+}\pi^{-}$ invariant mass distribution at $\sqrt{s} = 0.5,
1.96, 14$ TeV integrated over the full phase space (left panels) and
with the detector limitations in $\eta_{\pi}$ (right panels).
Results for the $\pi\pi$ continuum with the meson propagator
and with the cut-off parameters $\Lambda_{off}^{2}$ = 1.6, 2 GeV$^{2}$ 
(lower and upper dashed lines, respectively) as well as
with the generalized pion propagator and $\pi\pi$-rescattering (solid line) are presented.
We use GRV94 NLO (dotted lines) and GJR08 NLO (filled areas) collinear
gluon distributions. 
The absorption effects both for the signal and background were included in the calculations.}
\end{figure}

In Fig.~\ref{fig:dsig_dmpipi} we compare differential distributions of pions from the
$\chi_{c0}$ decay (see the peak at $M_{\pi\pi} \simeq 3.4$ GeV) with those for the continuum pions.
While left panels show the cross section integrated over the
full phase space, the right panels show results
including the relevant pion pseudorapidities restrictions 
$-1 < \eta_{\pi^{+}},\eta_{\pi^{-}} < 1$ (RHIC and Tevatron) and 
$-2.5 < \eta_{\pi^{+}},\eta_{\pi^{-}} < 2.5$ (LHC).
The $\chi_{c0}$ contribution is calculated with 
GRV94 NLO \cite{GRV} and GJR08 NLO \cite{GJR} collinear gluon distributions.

\begin{figure}[!ht]
\includegraphics[width = 0.23\textwidth]{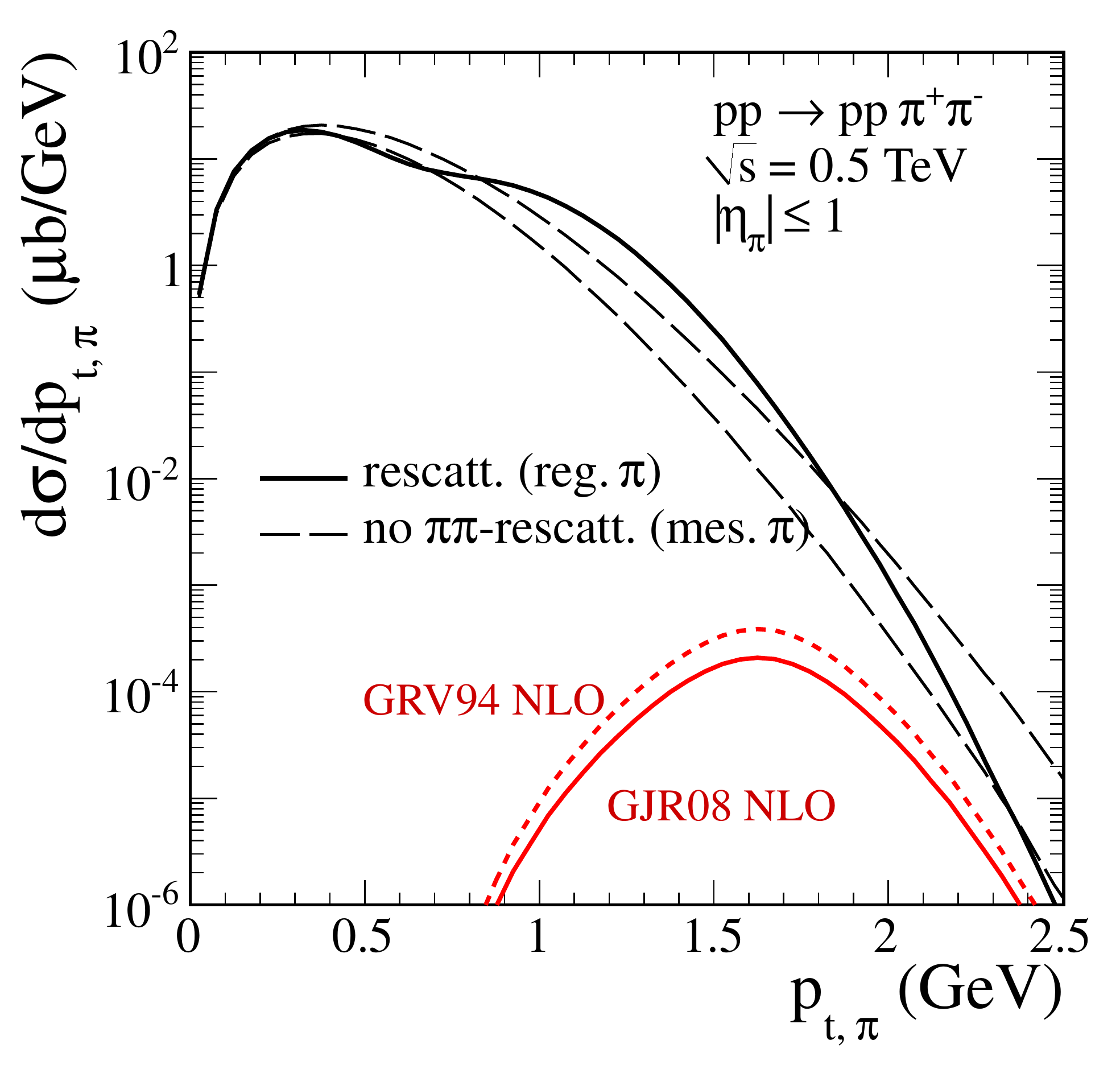}
\includegraphics[width = 0.23\textwidth]{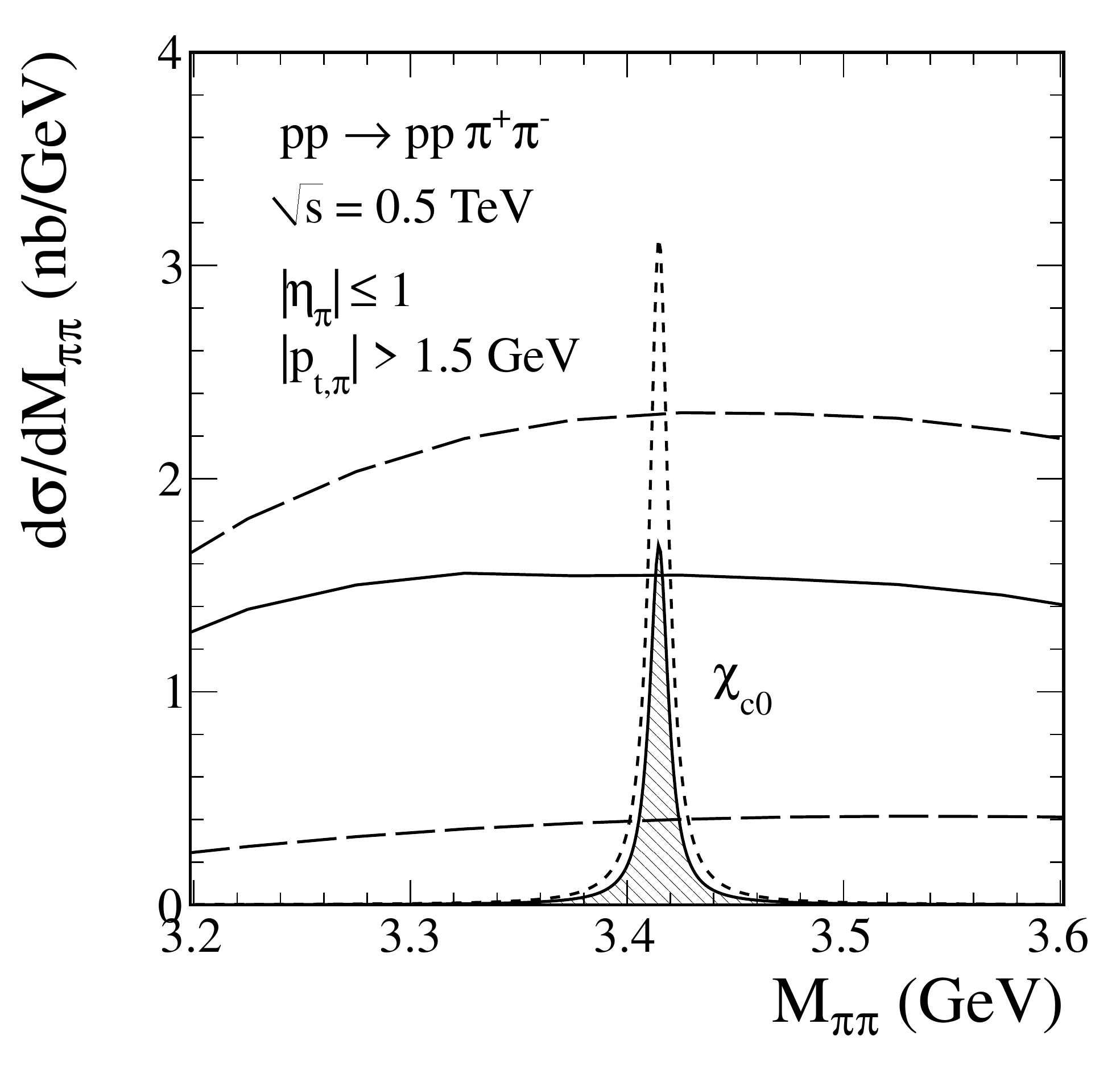}
\includegraphics[width = 0.23\textwidth]{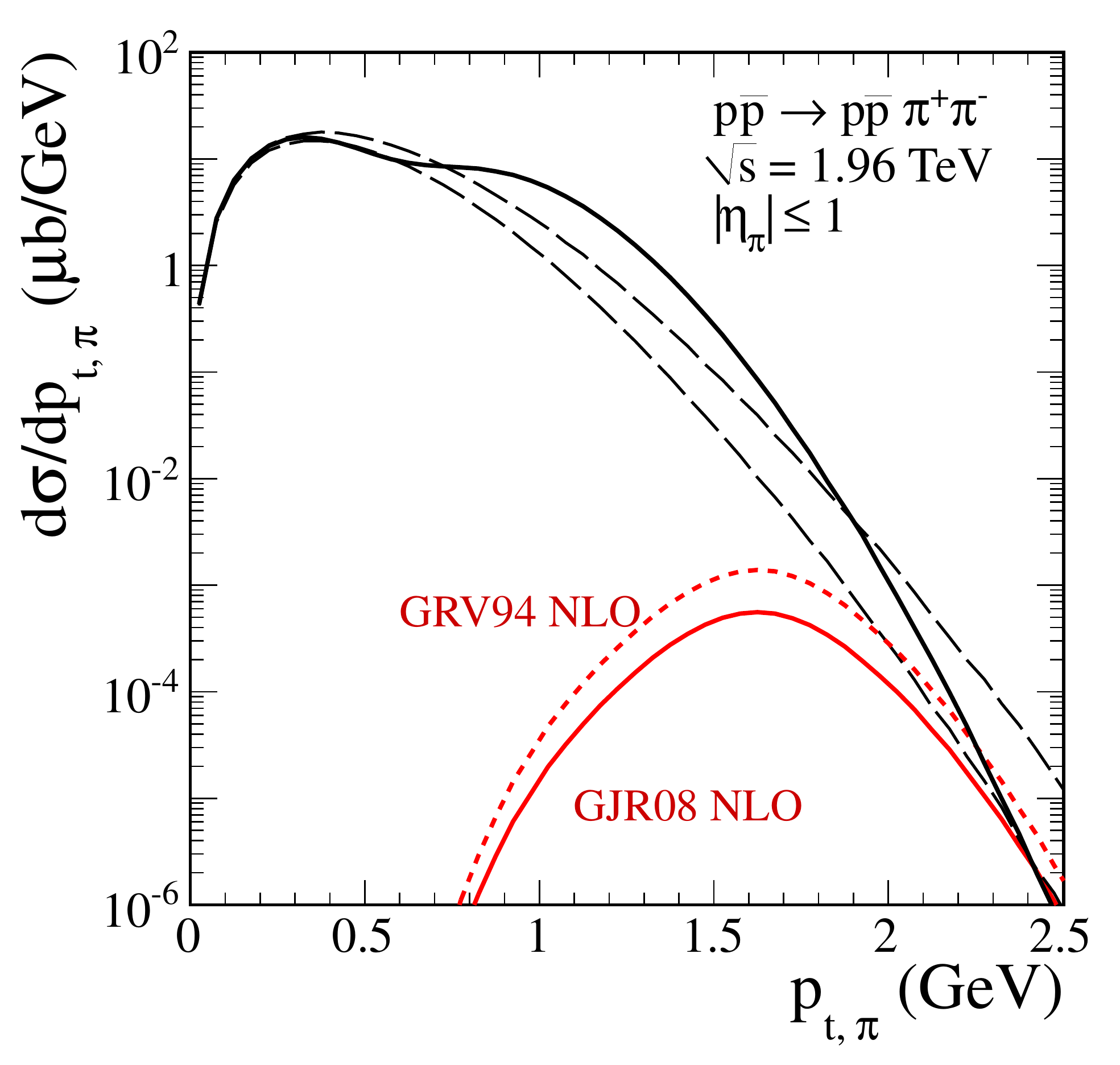}
\includegraphics[width = 0.23\textwidth]{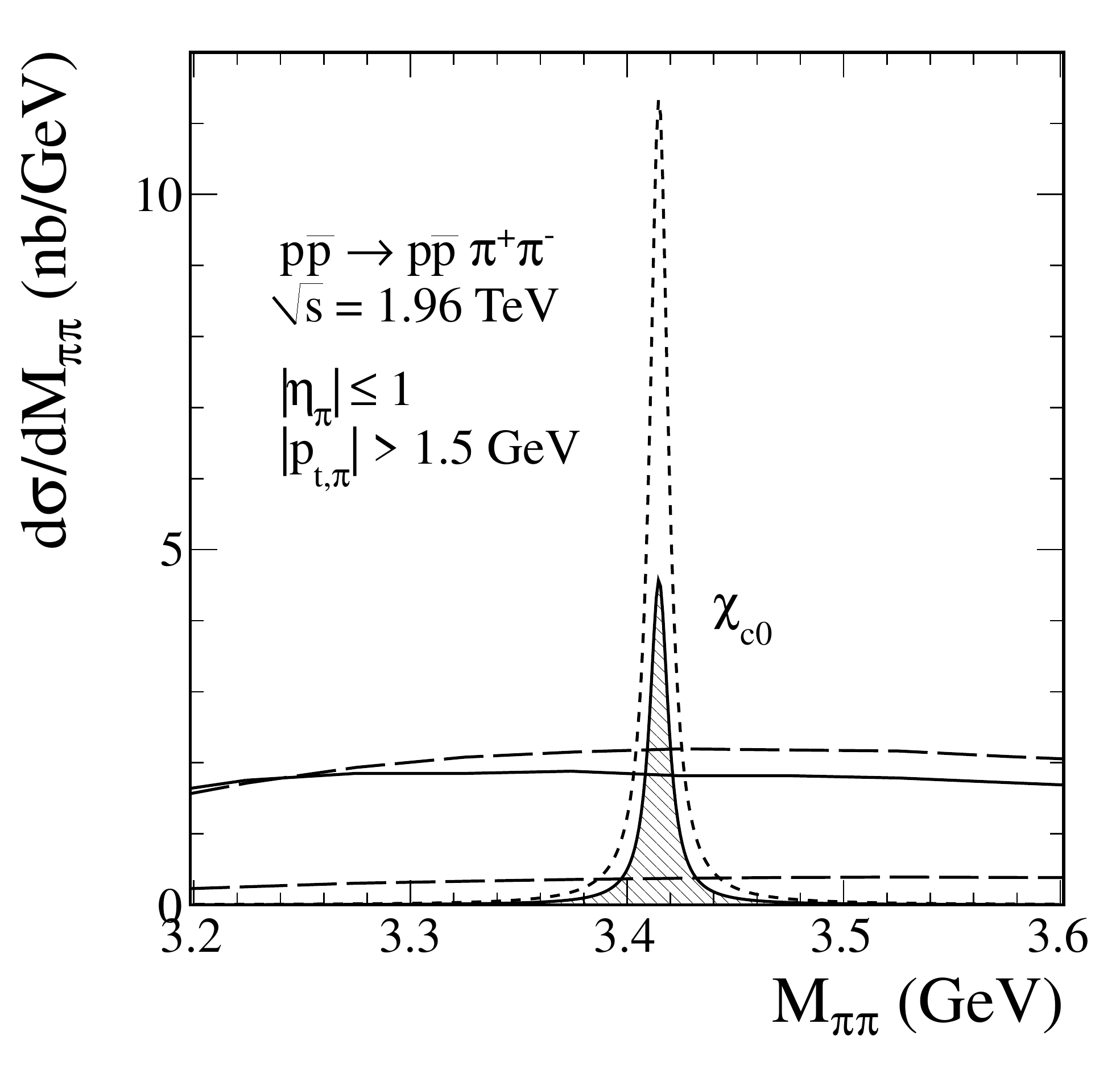}
\includegraphics[width = 0.23\textwidth]{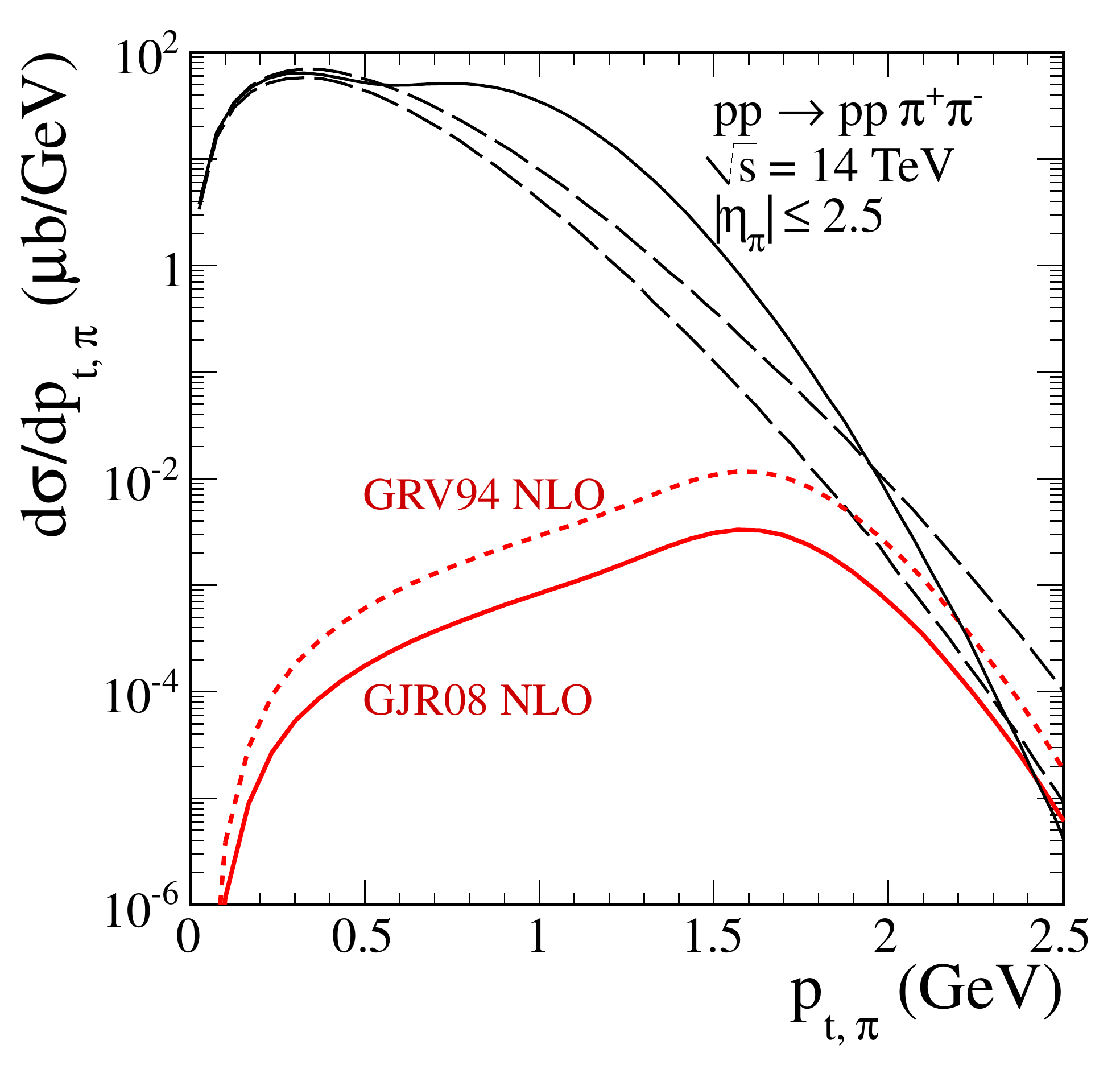}
\includegraphics[width = 0.23\textwidth]{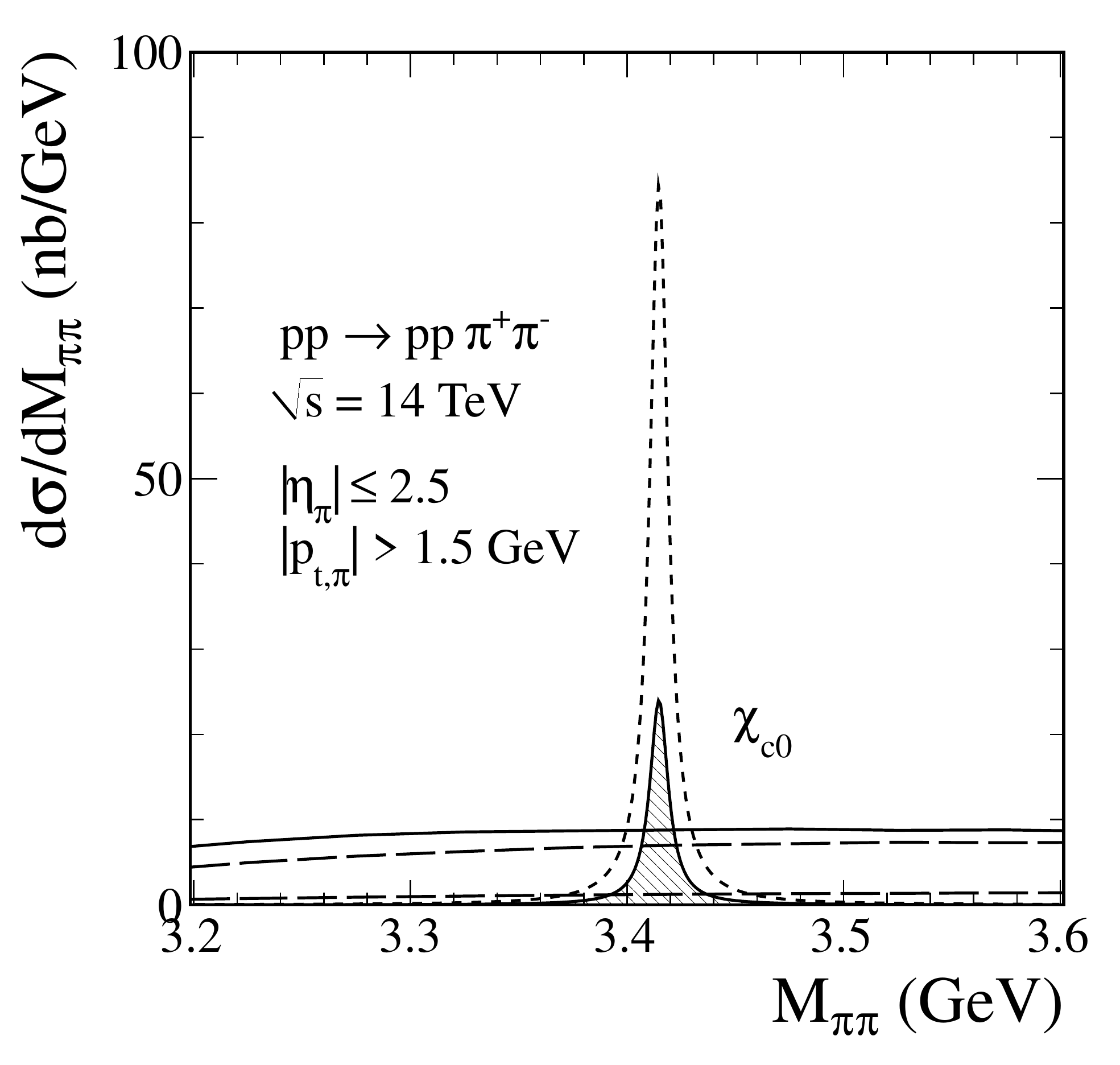}
  \caption{\label{fig:pt_lhc}
  \small
Left panels: Differential cross section $d\sigma/dp_{t,\pi}$ at $\sqrt{s} = 0.5,
1.96, 14$ TeV with cuts on the pion pseudorapidities.
The absorption effects both for the signal and background were included in the calculations.
Right panels: The $\pi^{+}\pi^{-}$ invariant mass distribution
with the relevant restrictions in the pion
pseudorapidities and pion transverse momenta.}
\end{figure}

In Fig.~\ref{fig:pt_lhc} we show
distributions in pion transverse momenta (left panels).
The pions from the $\chi_{c0}$ decay are placed at slightly larger transverse momenta. 
This can be therefore used to get rid of the bulk of the continuum by imposing
an extra cut on the pion transverse momenta.
In the right panels we show two-pion invariant mass distributions
with additional cuts on both pion transverse momenta
$|p_{t,\pi}| > 1.5$ GeV. Now the signal-to-background ratio is
somewhat improved especially at the Tevatron and LHC energies. 

\begin{figure}[!ht]
\includegraphics[width = 0.235\textwidth]{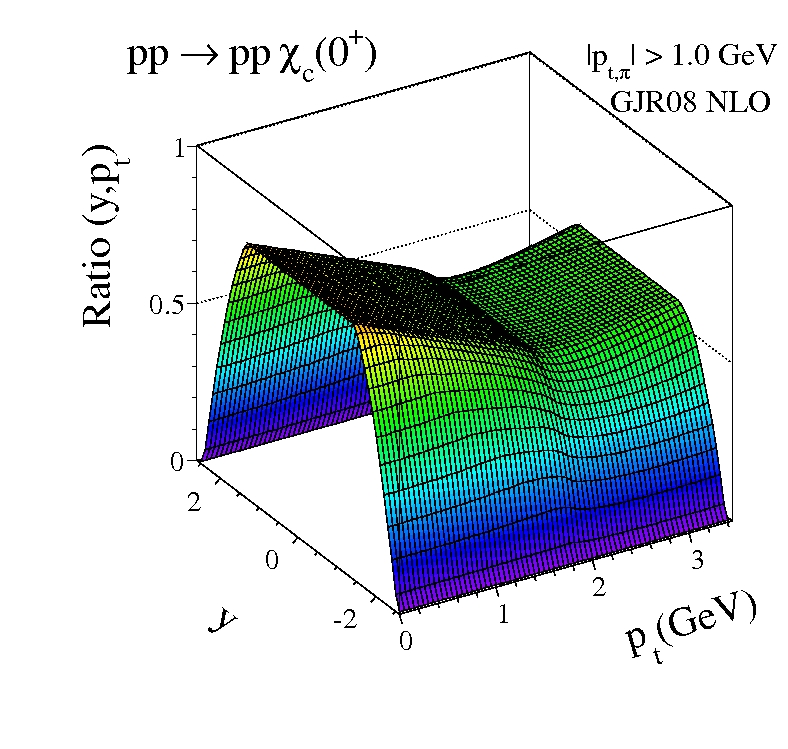}
\includegraphics[width = 0.235\textwidth]{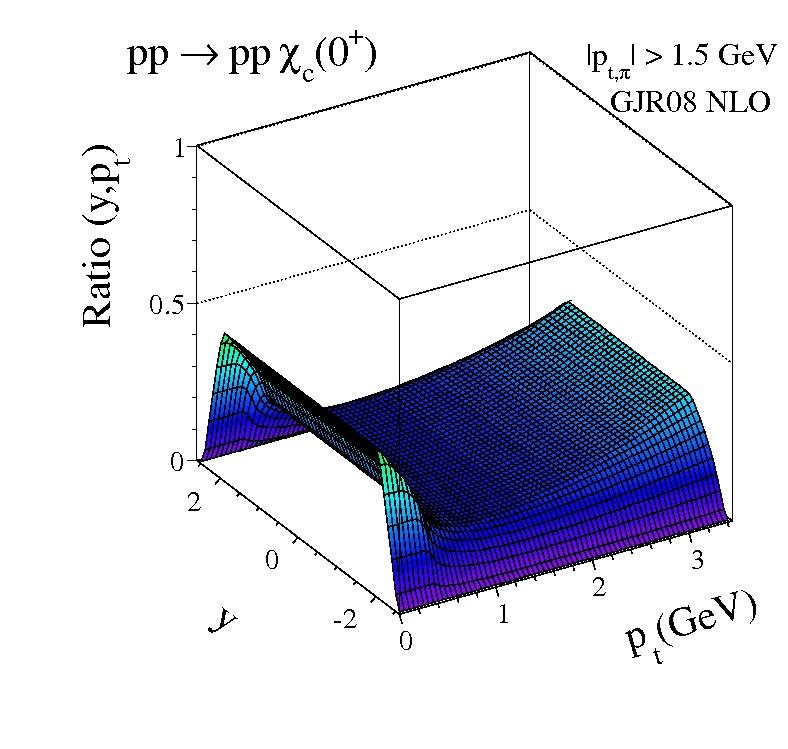}
  \caption{\label{fig:ratio}
  \small
Ratio of the two-dimensional cross sections in $(y,p_{t})$ for the
$pp \to pp \chi_{c0}$ reaction
with the relevant limitations on the pion pseudorapidities 
for the ATLAS or CMS detectors ($|\eta_{\pi}| < 2.5$) and a cuts on the pion transverse momenta $p_{t,\pi}$.}
\end{figure}

The main experimental task is to measure the distributions in the
$\chi_{c0}$ rapidity and transverse momentum. 
In Fig.~\ref{fig:ratio} we show the two-dimensional ratio
of the cross sections for the $\chi_{c0}$ meson in its rapidity and
transverse momentum:
\begin{eqnarray}\nonumber
\mathrm{Ratio}(y,p_{t}) =
\frac{d\sigma^{pp \to pp \chi_{c0}(\to \pi^{+} \pi^{-})}_{\mathrm{with \; cuts}} / dydp_{t}}
{d\sigma^{pp \to pp \chi_{c0}} / dydp_{t}}\,.
\label{ratio_formula}
\end{eqnarray}
The numerator includes limitations on $\eta_{\pi}$ and $p_{t,\pi}$.
These distributions provide a fairly precise evaluation of the expected
acceptances when experimental cuts are imposed.
The experimental data could be corrected by our two-dimensional
acceptance function to recover the distributions of interest.

\section{Conclusions}
\label{section:IV}

It was realized recently that the measurement
of exclusive production of $\chi_{c}$ via decay in the
$J/\psi + \gamma$ channel cannot give production cross sections for
different species of $\chi_{c}$.
In this decay channel the contributions of $\chi_{c}$ mesons
with different spins are similar and experimental resolution is not
sufficient to distinguish them.

We have analyzed a possibility to measure the
exclusive production of $\chi_{c0}$ meson in the proton-(anti)proton
collisions at the LHC, Tevatron and RHIC via $\chi_{c0} \to
\pi^{+}\pi^{-}$ decay channel. 
Since the cross section for
exclusive $\chi_{c0}$ production is much larger than that for
$\chi_{c1}$ and $\chi_{c2}$ and the branching fraction to the $\pi \pi$ channel
for $\chi_{c0}$ is larger than that for $\chi_{c2}$ ($\chi_{c1}$
does not decay into two pions) the two-pion channel should provide
an useful information about the $\chi_{c0}$ CEP.

We have performed detailed studies of several differential
distributions and demonstrated how to impose extra cuts in order to
improve the signal-to-background ratio. The two-pion background was
calculated in a simple model with parameters adjusted to low energy
data (see \cite{LS10, LPS11}). 
We have shown that relevant measurements at Tevatron and LHC
are possible. At RHIC the signal-to-background ratio is much worse
but measurements should be possible as well. Imposing cuts distorts
the original distributions for $\chi_{c0}$ in rapidity and
transverse momentum. We have demonstrated how to recover the
original distributions and presented the correction functions for
some typical experimental situations.






\end{document}